\documentclass[a4paper,11pt]{article}
\pdfoutput=1 
\usepackage{jheppub} 
\usepackage{subfigure}
\usepackage{amsmath}
\usepackage{braket}
\usepackage[T1]{fontenc} 
\usepackage{lineno}
\usepackage{caption}

\def\GeV{\ifmmode {\mathrm{\ Ge\kern -0.1em V}}\else \textrm{Ge\kern -0.1em V}\fi}%

\title{\boldmath Determination of the Transverse Momentum of W Bosons in Hadronic Collisions via Forward Folding Techniques}

\author[a]{Jakub Cuth}
\author[a,b]{Kyrylo Merkotan}
\author[a]{Matthias Schott}
\author[a]{Samuel Webb}
\affiliation[a]{Johannes Gutenberg-University, Mainz, Germany}
\affiliation[b]{Odessa National Polytechnic University, Odessa, Ukraine}

\emailAdd{mschott@cern.ch}

\abstract{
The measurement of the transverse momentum of W bosons in hadron collisions provides not only an important test of QCD calculations, but also is an important input for the precision measurement of the W boson mass. While the measurement of the Z boson transverse momentum is experimentally well under control, the available unfolding techniques for the W boson final states lead generically to relatively large uncertainties. In this paper, we present a new methodology to estimate the W boson transverse momentum spectrum, significantly improving the systematic uncertainties of current approaches.
}

\begin{document} 
\maketitle
\flushbottom
\newpage

\section{Introduction}	

At leading order, the electroweak vector bosons are produced in hadron collisions with zero momentum transverse to the beam line. A non-zero transverse momentum, $p_T$, is generated through the emission of partons in the initial state and therefore its measurement provides an important test of quantum chromo dynamic (QCD) calculations. Different approaches for the theoretical prediction of the $p_T$ spectra have been developed in recent years, using fixed-order calculations, parton shower models or resummed calculations, aiming for an accurate prediction of the low-$p_T$ and high-$p_T$ parts of the spectrum. 

The W boson production in hadron collisions is usually studied in its leptonic decay channels $W^\pm\rightarrow l^\pm \nu$ ($l=e,\mu$) due to the enormous di-jet background in the hadronic decay channel. The fundamental challenge in the measurement of the transverse momentum spectrum of the W boson, $p_T(W)$, is the missing information due to the decay neutrino. In contrast to the study of the Z boson transverse momentum, where both decay leptons can be reconstructed and therefore the transverse momentum of the boson can be calculated, this is not possible in the case of the W boson. 

Even though the underlying dynamics of the vector boson production are similar, the uncertainties on the $p_T(W)$ and $p_T(Z)$ measurements differ and are uncorrelated to a large extent. Even more importantly, the $p_T(W)$ spectrum has a direct impact on the measurement of the W boson mass at hadron colliders and hence should be tested experimentally. The available measurements (e.g. \cite{Aad:2011fp}),  provide only differential cross-sections for a few bins up to a $p_T(W)\approx 50\,\GeV$ with a relative precision of $3-5\%$, mainly limited by the unfolding procedure. In this paper, we present a new methodology that allows the extraction of a continuous spectrum with a significantly smaller associated methodology uncertainty.

The paper is structured as follows: A detailed explanation of the general methodology is given in Section \ref{sec:Met}, which is based on a functional description of the $p_T(W)$ spectrum. One possible example for such a functional description is given in Section \ref{sec:Func}, where its limitations are also discussed. A comparison of the traditional unfolding approach for the measurement of the $p_T(W)$ spectrum to the newly developed approach is presented in Section \ref{sec:comp}. The paper ends with a short summary and a conclusion in Section \ref{sec:Sum}.

\section{\label{sec:Met}Methodology}

\begin{figure}[tb]
\begin{minipage}[t]{7.5cm}
	\centering
	\includegraphics[width=7.4cm]{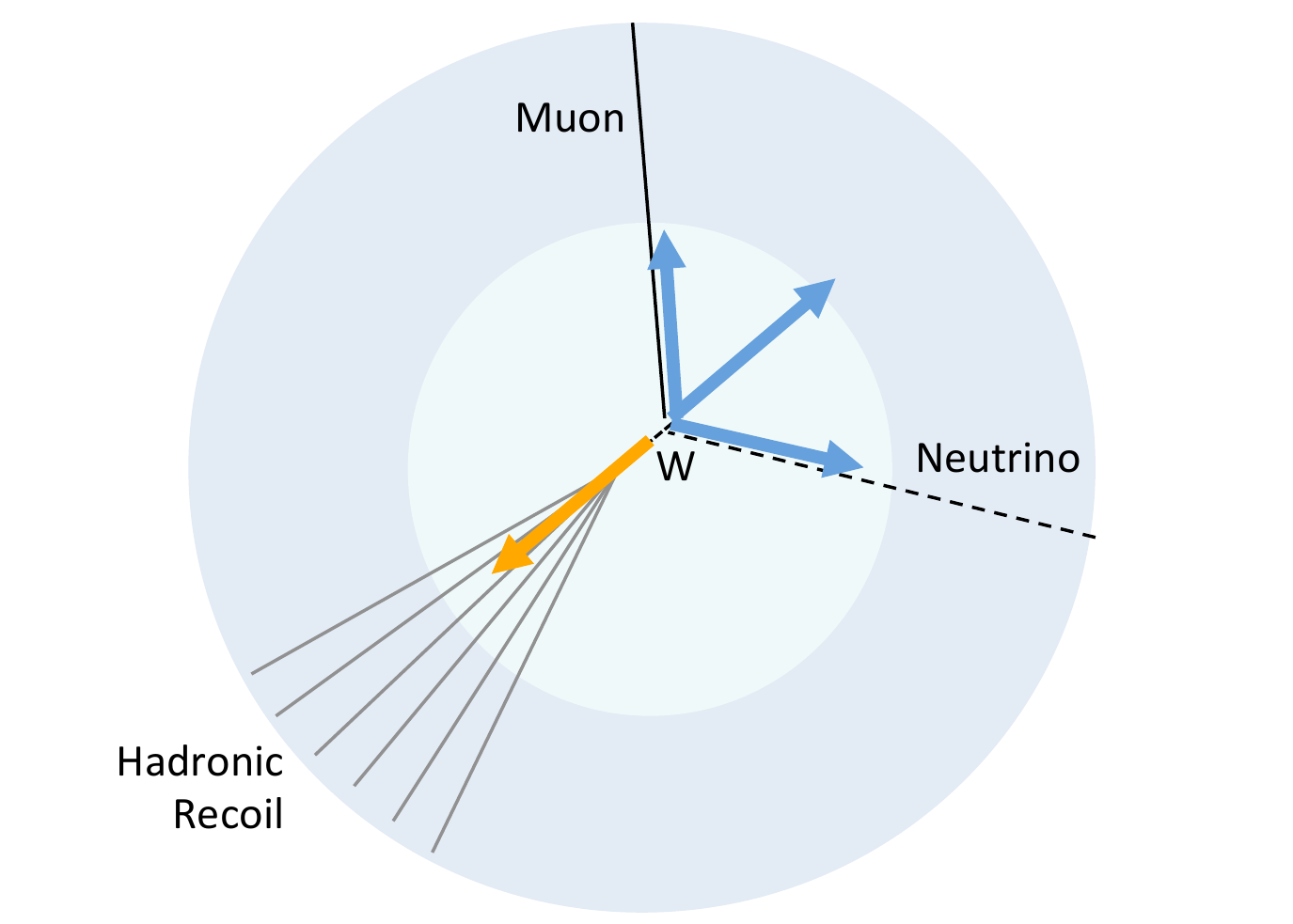}
	\caption{Schematic illustration of a leptonic W boson decay that is balanced by hadronic recoil.}
	\label{Fig:HadronicRecoil}
\end{minipage}
\hspace{0.1cm}
\hfill
\begin{minipage}[t]{7.5cm}
	\centering
	\includegraphics[width=7.4cm]{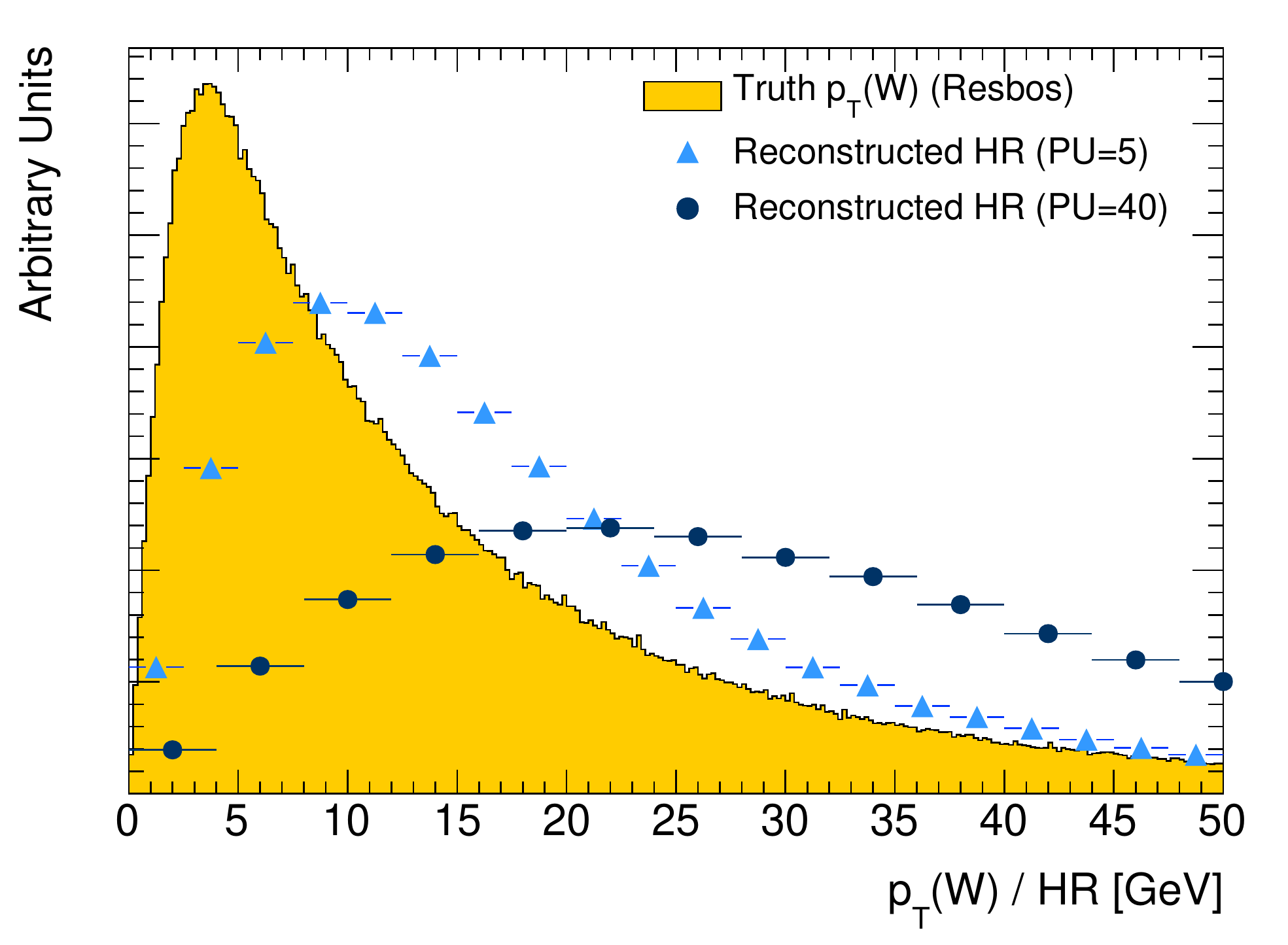}
	\caption{Reconstructed hadronic recoil based on a \textsc{Delphes} simulation of an LHC detector and truth $p_T(W)$ spectrum predicted by \textsc{Pythia8} at $\sqrt{s}=8\,$TeV with $\braket{\mu}=5$ and $\braket{\mu}=40$. Further details in Section \ref{sec:Events}.}
	\label{Fig:Problem}
\end{minipage}
\end{figure}

As discussed in the previous section, the $p_T(W)$ cannot be directly measured by the measurement of decay lepton kinematics. Therefore, the measurement of $\vec p_T(W)$ relies on the so-called hadronic recoil, $\vec u$. The hadronic recoil is defined per event as the vectorial sum of all measured transverse energies in the calorimeters except for the contribution from the decay leptons (Figure \ref{Fig:HadronicRecoil}). It can be interpreted as the energy stemming from the initial state radiation in the process $pp\rightarrow W + X$, which gives rise to $p_T(W)$. With an ideal detector, the vectorial sum of the measured hadronic recoil and the transverse momentum of the vector boson in the process must vanish, i.e. $\vec u + \vec p_T(W) = 0$. The fundamental quantity for the measurement of $p_T(W)$ is therefore the hadronic recoil. Experimentally, $\vec u$ is a complex quantity with a poor energy resolution since it is based on a large number of low energetic calorimetric measurements and depends strongly on the number of pile-up (PU) collisions $\braket{\mu}$ \footnote{The term pile-up denotes the number of hadron hadron interactions in the same bunch-crossing}. A typical energy resolution of $\vec u$ for a W boson with a transverse momentum of 40 GeV at an LHC detector is $\sim20\%$ for $\braket{\mu}=5$ and can rise to $\approx 35\%$ and $\approx 45\%$ for $\braket{\mu}=20$ and $\braket{\mu}=40$, respectively.

The experimental challenge is to infer information about the true $p_T(W)$ distribution from the measured distribution of $|\vec u|$, as illustrated in Figure \ref{Fig:Problem}. This  can usually be achieved by applying unfolding techniques, such as a Bayesian unfolding approach \cite{D'Agostini:1994zf}. This technique is used to unfold the reconstructed distribution of $|\vec u|$ to the true distribution $p_T(W)$ for selected events, taking into account bin-to-bin migration effects via a response matrix describing the probabilistic mapping from $p_T(W)$ to $|\vec u|$. This approach is typically chosen by high energy physics experiments to unfold kinematic distributions, e.g. \cite{Aad:2014xaa}. It is important to note that the chosen binning of the response matrix is crucial for the stability of the unfolded results. A basic quantity here is the stability $p$, which defines the probability of an event to be reconstructed in the same bin as it was generated. Typically stability values above 60\% lead to stable results of the unfolding procedure. Given the limited energy resolution of the hadronic recoil, rather large bin sizes have to be chosen for the unfolding of a $p_T(W)$ spectrum. Examples are discussed in more detail in Section \ref{sec:ExampleUnfolding}.

Our ansatz for measuring the $p_T(W)$ distribution is based on a forward folding approach using a functional parametrization of the distribution itself, which we denote as $f_{V}(p_T(V), a_1, a_2, ..., a_n)$. A potential choice of $f_V$ is discussed in detail in Section \ref{sec:Func}. For the following discussion it can be assumed that $f_V$ is flexible enough to describe any realistic transverse momentum distribution of vector bosons $V=W,Z$ in hadron collisions using only up to three functional parameters $a_i$ and returns the probability for a given transverse momentum to be produced. The goal of measuring the truth $p_T(W)$ distribution is now reduced to the determination of the functional parameters $a_i$, using the measured $|\vec u|$ distribution. The starting point for the evaluation of the functional parameters is the predicted distribution of $H(p^{MC}_T(W))$ by a Monte Carlo Event generator, the corresponding reconstructed hadronic recoil distribution after the full detector simulation $H(u^{MC})$ and the actual measured hadronic recoil distribution $H(u^{Data})$, where $H$ denotes the representation in form of a normalized histogram. In a second step, the function $f_V$ is fitted to $H(p^{MC}_T(W))$, resulting in initial parameters $a_i$. For each variation of the parameters $a_i$, the histogram $H(p^{MC}_T(W))$ and therefore also $H(u^{MC})$ can be reweighted on an event-by-event basis, such that $H(p^{MC}_T(W))$ corresponds to the new values of $a_i$. This reweighting procedure therefore transfers the possible variations of the underlying truth spectra to the observable $H(u^{MC})$. The difference between the predicted $H(u^{MC})$ and the observed distribution $H(u^{Data})$ can be quantified by a simple $\chi^2$ definition, i.e.

\begin{equation}
\chi^2 = \sum _i \frac{(H_i(u^{MC}) - H_i(u^{Data}))^2}{\sigma_i(u^{MC})^2 + \sigma_i(u^{Data})^2}
\end{equation}
where $H_i$ denotes the value of histogram bin $i$ and $\sigma_i$ is its corresponding uncertainty. By using a multi-parameter minimization algorithm, such a \textsc{Minuit} \cite{James:1975dr}, the parameters $a_i$ can be fitted such that the $\chi^2$ value (as difference between prediction and observation) is minimized. In this case we assume that the function $f_V$, corresponding to the fitted parameters $a_i$, describes the true spectrum of the vector boson. 

The advantage of this procedure lies in the fact, that the expected functional form of the transverse momentum distribution reduces the measurement problem to the determination of a few functional parameters by taking into account a detailed observed kinematic spectrum. This bypasses the restrictions due to bin purities as described in the traditional strategy. Clearly, this methodology is limited to which extent the function form $f_V$ is flexible enough to reassemble all possible realistic $p_T(W)$ distributions. These limitations and the actual functional form of $f_V$ are discussed in the next section.

\section{\label{sec:Func}Semi-empirical Description of the Vector Boson Transverse Momentum Spectra}

\subsection{Functional Description}

Modern event generators that have specialized in the prediction of the vector boson transverse momentum in hadronic collisions, such as \textsc{Resbos} \cite{Ladinsky:1993zn}, \cite{Balazs:1997xd}, \cite{Landry:2002ix}, rely on resummed calculations at next-to-next-to-leading order (NNLO) and next-to-next-to-leading-log approximations (NNLL). The NNLO corrections dominate the high $p_T$ part of the vector boson spectrum, while the resummed NNLL calculations aim for a good description of the low $p_T$ part. These event generators typically provide a very accurate description of the measured Z boson transverse momenta, once their model parameters have been fitted to the available measurements. In principle, these event generators could serve as basis for a functional description $f_{V}$, as discussed in Section \ref{sec:Met}. However, this would imply regenerating an enormous number of events for each fitting step, which leads to computational times that are currently not yet reachable. 

Therefore, we propose a semi-empirical parametrization that relies on a number of simple arguments, first introduced in \cite{Boonekamp:2010ik}. Consider a vector boson production at high energy, and at given mass and rapidity, so that the parton momentum fractions at the hard vertex are small and fixed. In the low transverse momentum region, the repeated gluon emission in the initial state generates a Gaussian transverse momentum distribution. Along the $x$ and $y$ axes, this \textit{random walk} leads to a distribution proportional to $$f(p_{x,y};\sigma)  \,\, dp_{x,y} \sim e^{-\frac{p^2_{x,y}}{2\sigma^2}} \,\, dp_{x,y}.$$ 

The $\sigma$ parameter represents the spread of the $p_{x,y}$ distribution after all emissions and, in a naive picture, could be seen as representing the average number of emitted gluons times their average transverse momentum: $\sigma \sim \sqrt{N_g} \times p_{T,g}$. Moving to polar coordinates, the distribution becomes:    

\begin{eqnarray}
\label{eqnparaorg}
f(p_x;\sigma)  \,\, f(p_y;\sigma)  \,\, d p_{x}  \,\, d p_{y} &\sim& e^{-\frac{p^2_{x}}{2\sigma^2}} \,\, e^{-\frac{p^2_{y}}{2\sigma^2}} \,\, d p_x \,\, d p_y = \nonumber \\
  &=&  e^{-\frac{p_T^2}{2\sigma^2}} \,\, p_T \,\, d p_T \,\,
  d \phi \,\, \equiv \,\, g_1(p_T;\sigma) \,\, d p_T \nonumber
\end{eqnarray}
after a trivial azimuthal integral. At higher $p_T$, the shape is dominated by a power law behavior representing the parton density functions (PDFs) and the perturbative matrix element: 

\[g_2(p_T;a) \sim 1/p_T^\alpha.\]
The transition between the two descriptions is controlled by a parameter $n$, defined such that $p_T^{match} =n \times \sigma$ and satisfies smoothness conditions (the function and its derivative are continuous). The complete parametrization is, ignoring an overall normalization factor:

\begin{equation}
\label{eqnpara}
g(p_T; \sigma, \alpha, n) = 
 \begin{cases} 
      p_T \cdot e^{-\frac{p_T^2}{2\sigma^2}} 														& p_T \leq n \cdot \sigma \\
      p_T \cdot \frac{(\frac{\alpha}{n})^\alpha \, e^{-n^2/2}}{(\frac{\alpha}{n} - n+ \frac{p_T}{\sigma})^\alpha},		 & p_T > n \cdot \sigma 
   \end{cases}
\end{equation}
where the parameters $\alpha$, $n$, and $\sigma$ are all positive definite. 

\subsection{\label{Sec:accTests}Accuracy Tests and Improvements}

\begin{figure}[b]
\begin{minipage}[t]{7.5cm}
	\centering
	\includegraphics[width=7.4cm]{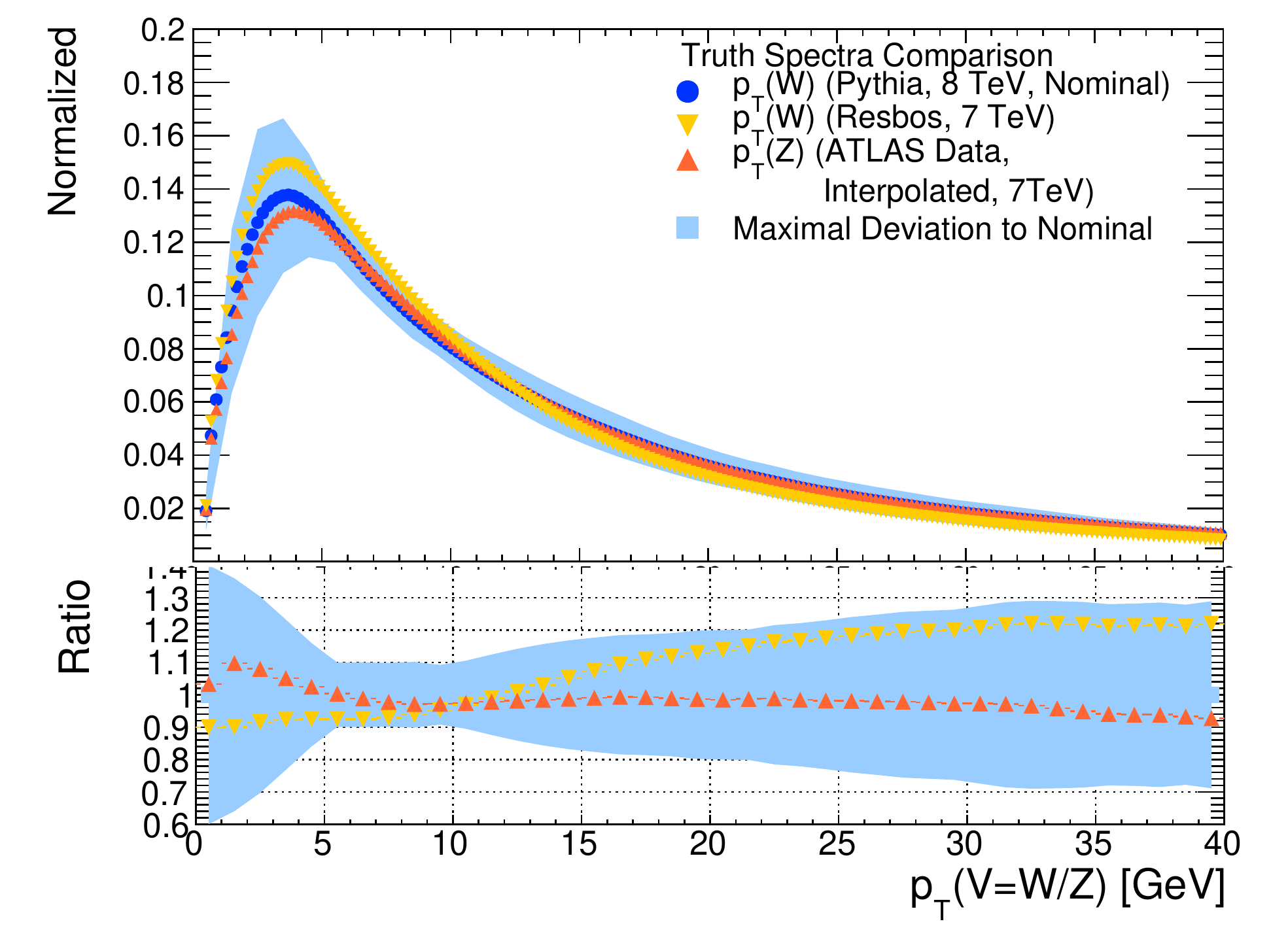}
	\caption{Predicted truth spectrum of $p_T(W)$ by \textsc{Pythia8}, normalized between $0$ and $40\,\GeV$ including the envelope of all other predicted spectra according to Tab. \ref{DataTable}.\vspace{0.8cm}}
	\label{Fig:AllTruthSpectra}
\end{minipage}
\hspace{0.1cm}
\hfill
\begin{minipage}[t]{7.5cm}
	\centering
	\includegraphics[width=7.4cm]{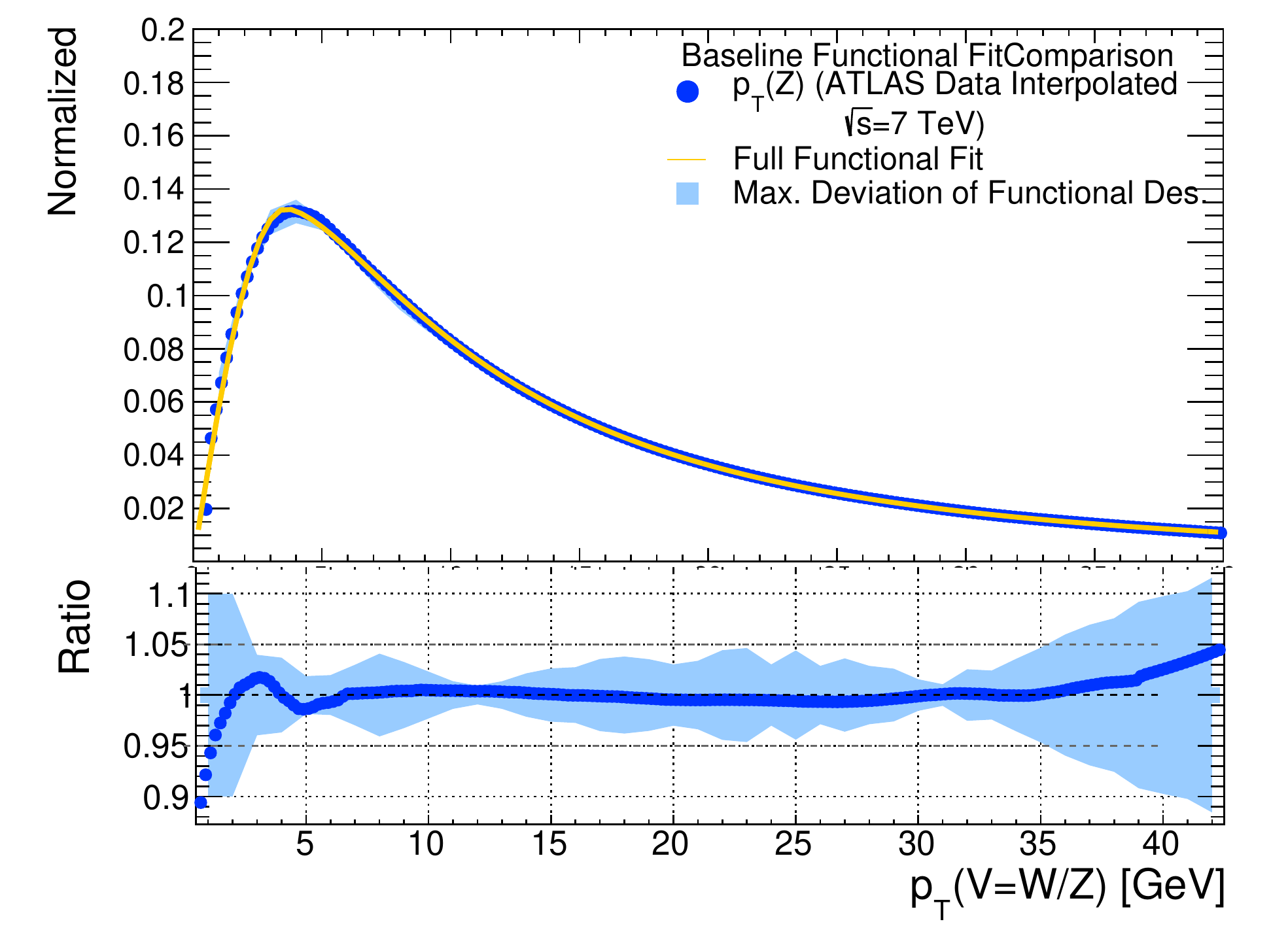}
	\caption{$p_T(Z)$ spectrum predicted by \textsc{Pythia8} at $\sqrt{s}=8$\,TeV with the fit of Function \ref{eqnparaorg}. The ratio shows the deviations between the fitted function and the data, as well as the accuracy band, determined by all available MC Sets from Table \ref{DataTable}.}
	\label{Fig:FuncBase}
\end{minipage}
\end{figure}

We have chosen to study a large variety of predicted $p_T(V=W/Z)$ spectra using different Monte Carlo event generators with different settings in order to test the accuracy of the proposed functional form. An overview of the samples is given in Table \ref{DataTable}. These predicted spectra of $p_T(V=W/Z)$ are based on different theoretical approximations and assumptions, vastly varied constants and scales. While the \textsc{Pythia8} \cite{Sjostrand:2007gs} predictions are based on a leading-order (LO) and parton shower (PS) calculation, \textsc{Powheg} \cite{Alioli:2010xd}, \cite{Frixione:2007vw} provides a next-to-leading order prediction and \textsc{Resbos} uses resummed calculations. A different parton shower model is provided by the \textsc{Sherpa} \cite{Gleisberg:2008ta} prediction. The predicted normalized $p_T(W)$ spectrum in 8 TeV proton-proton collisions is shown in Figure \ref{Fig:AllTruthSpectra}, where also the relative difference to all other predictions is visualized. The predictions vary by up to 40\% in the very low $p_T$ region and up to $30\%$ for transverse momentum values above 30 GeV. 

We assume in the following that the precision to which the functional description is able to describe these MC predictions is the same as for the true $p_T(V)$ distribution or in other words, that the true $p_T(V)$ spectrum is within the band shown in Figure \ref{Fig:AllTruthSpectra}. Hence we define the accuracy of a given function $f$, as the maximal deviation of the function from any of the MC predictions $h_i$ at a given transverse momentum value, i.e. $max(|f(p_T(V)) - h_i(p_T(V))|)$, where $i$ is the index of the given MC prediction. Hence the accuracy of a given functional form can be interpreted as the envelope of the differences to all available  predictions. This is shown in Figure \ref{Fig:FuncBase} for Equation \ref{eqnpara}, where we fit the measured $p_T(Z)$ distribution, provided by the ATLAS Experiment \cite{Aad:2014xaa} at $\sqrt{s}=7$\,TeV, and illustrate the accuracy envelope based on the MC samples of Table \ref{DataTable} in the ratio plot. Two things can be noted: First of all, the measured $p_T(Z)$ spectrum is far within the accuracy band of the functional description. This is not surprising, as the variety of MC predictions is very large, leading to a conservative estimate of the achievable accuracy. Secondly, even though the functional form given by Equation \ref{eqnpara} describes the basic features of the vector boson momentum distribution correctly, it has problems describing the rising edge and the peak region.

\begin{table}[htb]
\begin{center}
\small
\begin{tabular}{c|c|c|c|c|c}
\hline 
\hline 
MC Event Generator 		&	Process									& $\sqrt{s}$ [TeV]	& Settings							& PDF-Set 	& Events	\\
\hline 
\textsc{Pythia8} 			&	$pp\rightarrow W \rightarrow \mu\nu$\,			& 8 / 13			& Nominal						& CT10				& $10^7$		\\
						&	$pp\rightarrow Z \rightarrow \mu\nu$\,			& 8 				& 								& \cite{Gao:2013xoa}	& $10^7$			\\
\hline
\textsc{Pythia8} 			&	$pp\rightarrow W \rightarrow \mu\nu$\,			& 8 / 13			& Nominal						& MSTW				& $10^7$		\\
						&	$pp\rightarrow Z \rightarrow \mu\nu$\,			& 8 				& 								& \cite{Watt:2012tq}		& $10^7$			\\
\hline
\textsc{Pythia8} 			&	$pp\rightarrow W \rightarrow \mu\nu$\,			& 8 				& $\alpha_s=0.9\cdot\alpha_s^{nom}$	& CT10				&$10^7$			\\
\hline 
\textsc{Pythia8} 			&	$pp\rightarrow W \rightarrow \mu\nu$\,			& 8 				& $\alpha_s=1.1\cdot\alpha_s^{nom}$	& CT10				& $10^7$			\\
\hline 
\textsc{Pythia8} 			&	$pp\rightarrow W \rightarrow \mu\nu$\,			& 8 				& $\mu_F=2\cdot\mu_F^{nom}$		& CT10				& $10^7$			\\
						&											& 				& $\mu_R=2\cdot\mu_R^{nom}$		& 					& $10^7$			\\
\hline 
\textsc{Pythia8} 			&	$pp\rightarrow W \rightarrow \mu\nu$\,			& 8 				& $\mu_F=0.5\cdot\mu_F^{nom}$		& CT10				& $10^7$			\\
						&											& 				& $\mu_R=0.5\cdot\mu_R^{nom}$		& 					& $10^7$			\\
\hline 
\textsc{Sherpa} 			&	$pp\rightarrow Z \rightarrow \mu\mu$\,			& 8 				& Nominal						& CT10				& $10^7$			\\
\hline 
\textsc{PowHeg+Pythia} 		&	$pp\rightarrow Z \rightarrow \mu\mu$\,			& 8 				& Nominal						& CT10				& $10^7$			\\
(AU2-Tune)				&											&  				& 								& 					& $10^7$			\\
\hline 
\textsc{PowHeg+Pythia} 		&	$pp\rightarrow Z \rightarrow \mu\mu$\,			& 8 				& Nominal						& CT10				& $10^7$			\\
(AZ-Tune)					&											&  				& 								& 					& $10^7$			\\
\hline 
\textsc{ResBos} 			&	$pp\rightarrow W \rightarrow \mu\nu$\,			& 7 				& Nominal						& CT10				& $10^7$			\\
						&	$pp\rightarrow Z \rightarrow \mu\nu$\,			& 8 				& 								& 					& $10^7$			\\
\hline 
\hline 
\end{tabular}
\end{center}
\caption{Overview of 13 different MC predictions that are used to test the performance of the $p_T(V)$ measurement methodologies.} 
\label{DataTable}
\end{table}

\begin{figure}[tb]
\begin{minipage}[t]{7.5cm}
	\centering
	\includegraphics[width=7.4cm]{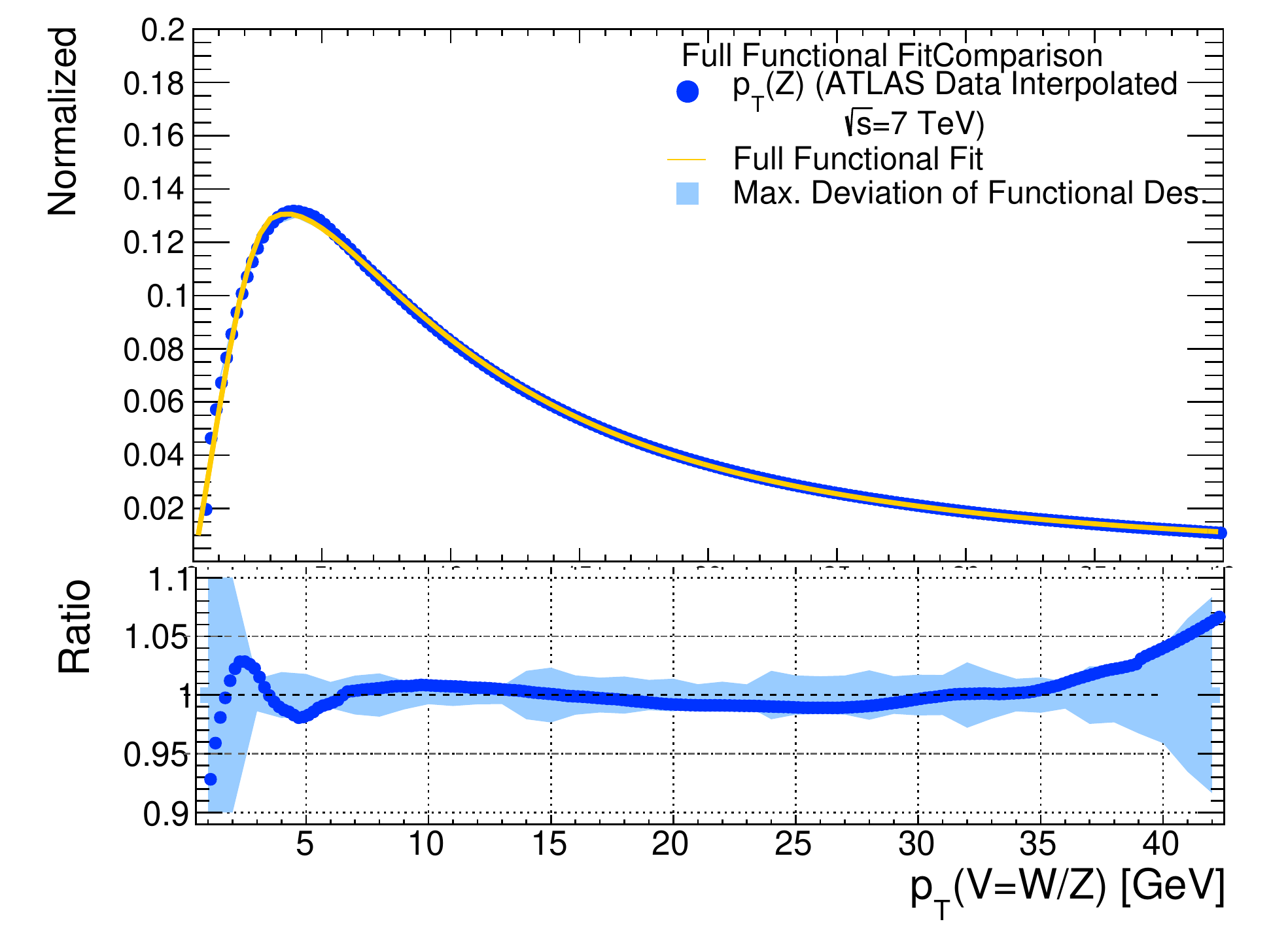}
	\caption{$p_T(Z)$ spectrum predicted by \textsc{Pythia8} at $\sqrt{s}=8$\,TeV with the fit of Function \ref{eqnpara}. The ratio shows the deviations between the fitted function and the data, as well as the accuracy band, determined by all available MC Sets from Table \ref{DataTable}.}
	\label{Fig:Func1}
\end{minipage}
\hfill
\hspace{0.1cm}
\begin{minipage}[t]{7.5cm}
	\centering
	\includegraphics[width=7.4cm]{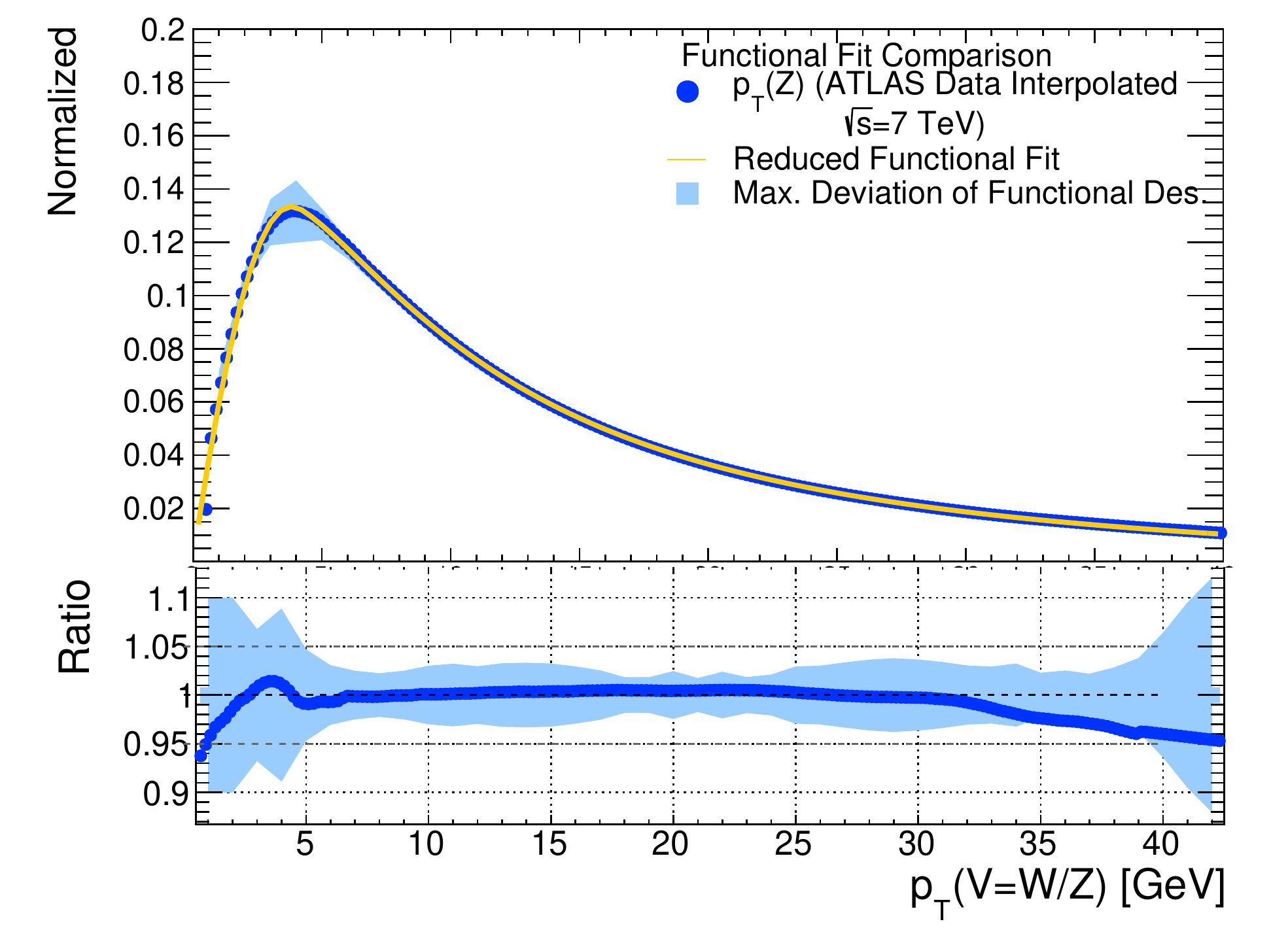}
	\caption{$p_T(Z)$ spectrum predicted by \textsc{Pythia8} at $\sqrt{s}=8$\,TeV with the fit of Function \ref{eqnpara2}. The ratio shows the deviations between the fitted function and the data, as well as the accuracy band, determined by all available MC Sets from Table \ref{DataTable}.}
	\label{Fig:Func2}
\end{minipage}
\end{figure}

Therefore, we modified function \ref{eqnpara} by adding two further parameters:

\begin{equation}
\label{eqnpara2}
g(p_T; \sigma, \alpha, n) = 
 \begin{cases} 
      (p_T+(p_T)^\rho) \cdot e^{-\frac{p_T^2}{2\sigma^2}} 																				& p_T \leq n \cdot \sigma \\
      (p_T+(p_T)^\rho) \cdot \frac{(\frac{\alpha}{n})^\alpha \, e^{-n^2/2}}{(\frac{\alpha}{n} - n+ \frac{p_T}{\sigma})^\alpha}+ \tau \cdot (p_T-\sigma\cdot n),		 & p_T > n \cdot \sigma 
   \end{cases}
\end{equation}
where the new parameter $\rho$ modifies the rising behavior via a power law, while the parameter $\tau$ modifies the falling part of the spectrum via an additional linear function. The corresponding accuracy envelope of this function with respect to the MC samples in Table \ref{DataTable} does not exceed a 2\% deviation anywhere between 2 and 35 \GeV\,of the vector boson transverse momentum. Fitting five parameters indirectly via a hadronic recoil spectrum leads to significant challenges during the minimization procedure\footnote{Further constraints could be given by hadronic recoil distributions in different rapidity bins of the charged decay lepton. The lepton $p_T$ spectrum also contains information about $p_T(W)$, but is correlated to $m_W$.}. However, it turned out that the two new parameters $\rho$ and $\tau$ are not independent from the first three parameters. In fact, the first additional parameter can be fixed to $\rho \approx 0.84$, while $\tau$ can be approximated by $\tau \approx (0.6\cdot \alpha^2 - 2.3\cdot \alpha - 1.1)/1000$, leading to a function with three free parameters. It should be noted, that the empirical fixation of $\rho$ and $\tau$ does not invalidate the usage of the function \ref{eqnpara2} for the fitting procedure. The resulting accuracy band of Equation \ref{eqnpara2} including the parameter fixation for all MC predictions is shown in Figure \ref{Fig:Func2}, again also indicating the measured $p_T(Z)$ spectrum. The uncertainty band does not exceed a 3\% deviation starting with $p_T$ larger than 5 GeV up to 40\,GeV. Again, the measured spectrum lies within the predicted accuracy band. The dominating limitations on the accuracy are due to the $\alpha_s$ variations of the \textsc{Pythia8} sample. We will use the parametrization of Equation \ref{eqnpara2} with fixed parameters $\rho$ and $\tau$ for all following studies.  

\section{\label{sec:comp}Comparison of Functional Forward Folding and Unfolding Techniques}

In this section, we will describe the actual determination of the $p_T(W)$ spectrum based on the measured hadronic recoil distribution $H(u^{Data})$. We compare the traditional (Bayesian) unfolding approach with the forward functional folding technique. The event and detector simulation, used in our analyses, is briefly described in Section \ref{sec:Events}, the results from the unfolding approaches are discussed in Section \ref{sec:ExampleUnfolding}, while the performance of the new approach is presented in Section \ref{sec:FunctionalFit}. Finally, we compare the results of both methods in Section \ref{sec:Comp}.

\subsection{\label{sec:Events}Event and Detector Simulation}

The baseline MC sample was chosen to be $pp\rightarrow W \rightarrow \mu\nu$ simulated using the baseline \textsc{Pythia8} with nominal settings. This sample was used as input to the \textsc{Delphes} framework \cite{Ovyn:2009tx}, using the ATLAS like detector settings to simulate a full detector response. In total $10^7$ detector simulated events were generated, using pile-up conditions of $\braket{\mu}=5$ and $40$, where $\braket{\mu}$ indicates the average number of interactions per bunch crossing, i.e. pile-up (PU) collisions. These samples are referred to as simulated MC samples in the following. The truth transverse momentum of the W boson and the expected measured hadronic recoil distribution by \textsc{Delphes} for all events was already shown in Figure \ref{Fig:HadronicRecoil}. As mentioned in Section \ref{sec:Met}, the sensitivity of the measured hadronic recoil $u$ on $p_T(W)$ significantly reduces for high pile-up environments. 

Since we want to study the pure impact of unfolding, we can assume a perfect understanding of our detector. Hence the simulated MC samples can be used to produce pseudo-data of measured hadronic recoil distributions for a variety of MC predictions. For this, we reweight the MC truth $p_T(W)$ distribution of the nominal \textsc{Pythia8} sample to each of the other MC predictions in Table \ref{DataTable}. This then also yields a reweighted distribution of the simulated measured hadronic recoil distribution, which is then used as pseudo-data. In order to test the performance of the unfolding techniques, we take this pseudo-data for a given MC generator as input, use the simulated MC samples for the detector description and test, if the original vector boson $p_T$ distribution of the given MC generator can be recovered.

\subsection{\label{sec:ExampleUnfolding}Expected Performance of an Unfolding Approach}

The most common unfolding technique used in high energy physics analyses, is the Bayesian unfolding approach \cite{D'Agostini:1994zf}. Hence it was chosen to test the performance of this approach for the measurement of the $p_T(W)$ spectrum based on the measured hadronic recoil distribution. Two key parameters, namely the purity and stability, determine the quality of the unfolding. The purity defines the probability that an observable, was generated at truth level in the same bin as it was finally reconstructed, i.e. in our case that the measured value of $u$ is in the same bin as the generated $p_T(W)$ value normalized by the number of reconstructed events in the given bin. The stability is similarly defined, but normalized by the number of truth events in the given bin. Usually it is assumed that the Bayesian unfolding technique leads to sufficiently stable results, when a purity and stability above 60\% is achieved. 

\begin{figure}[tb]
    \begin{center}
        \includegraphics[width=0.49\textwidth]{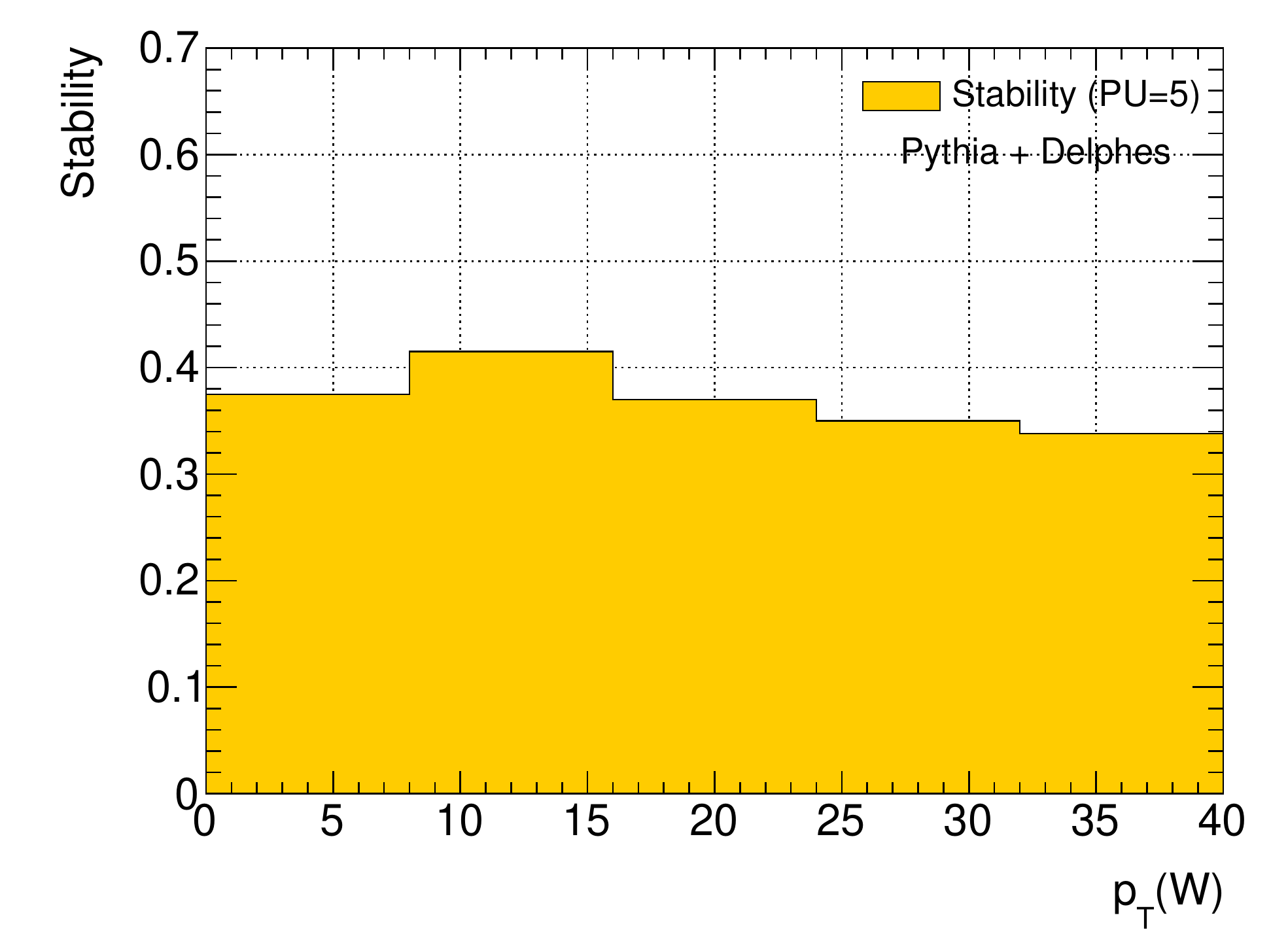}
        \includegraphics[width=0.49\textwidth]{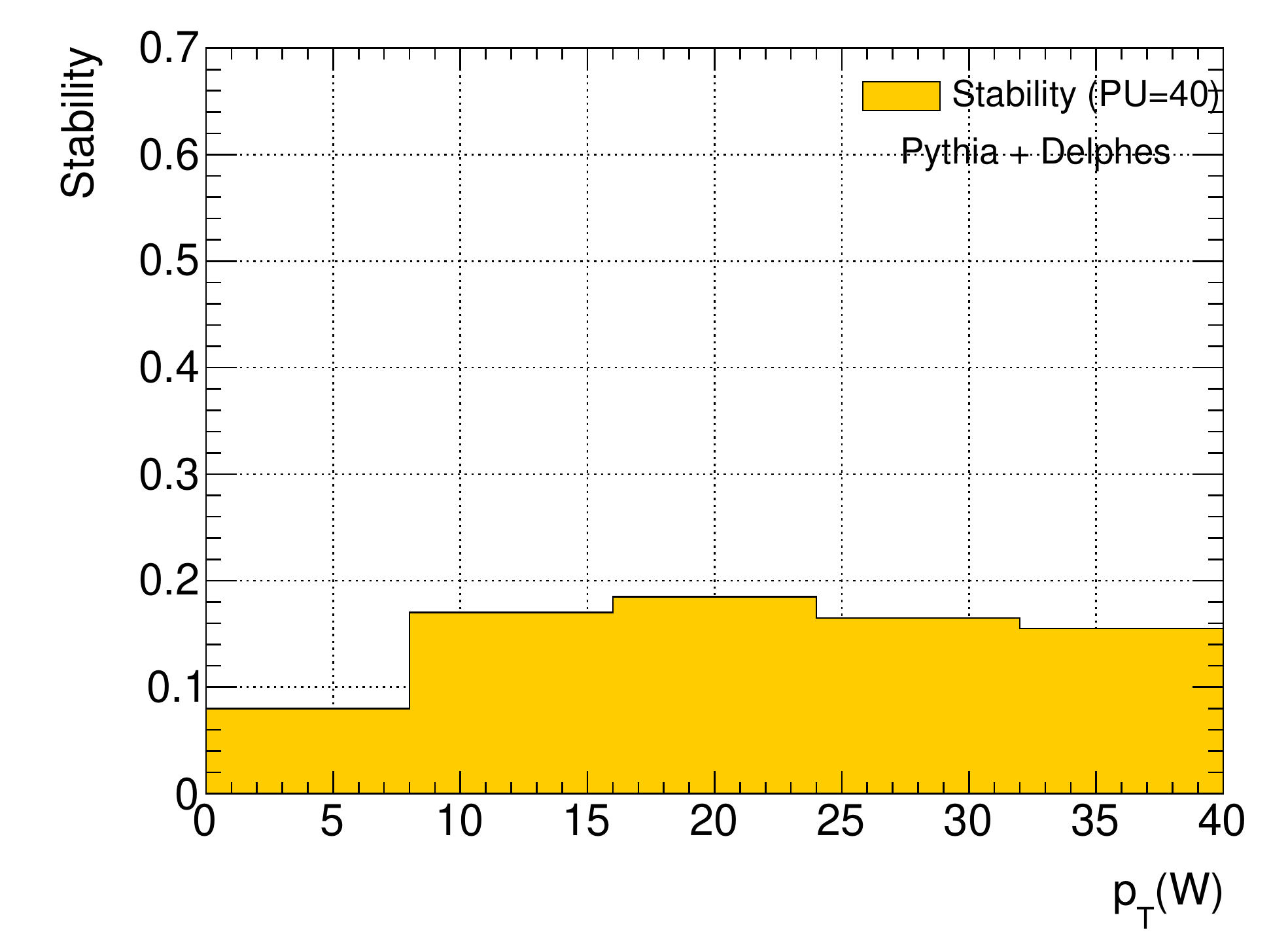}
        \caption{Expected stability distribution based on a DELPHES simulation of the ATLAS Detector at $\sqrt{s}=8\,$TeV with $\braket{\mu}=5$ and $\braket{\mu}=40$ for two different binnings.}
        \label{Fig:Purity}
    \end{center}
\end{figure}

Clearly, both the purity and stability depend largely on the reconstruction resolution of the given observable and hence also on the chosen binning. Larger bin sizes will lead to higher purities, but lose information on the functional shape of the underlying observable. In order to allow for a fair comparison, we have chosen an 8 GeV binning\footnote{Smaller binnings, e.g. by 1 GeV, led to extremely large uncertainties in the subsequent unfolding and hence have not been considered further.} for the Bayesian response matrix as it still contains limited knowledge on the basic functional shape of the $p_T(W)$ spectrum, even though the highly important peak region is covered by one bin. The response matrix was then filled by the simulated samples with $\braket{\mu}=5$ and $\braket{\mu}=40$. The corresponding stability values are shown in Figure \ref{Fig:Purity}.

\begin{figure}[tb]
    \begin{center}
        \includegraphics[width=0.49\textwidth]{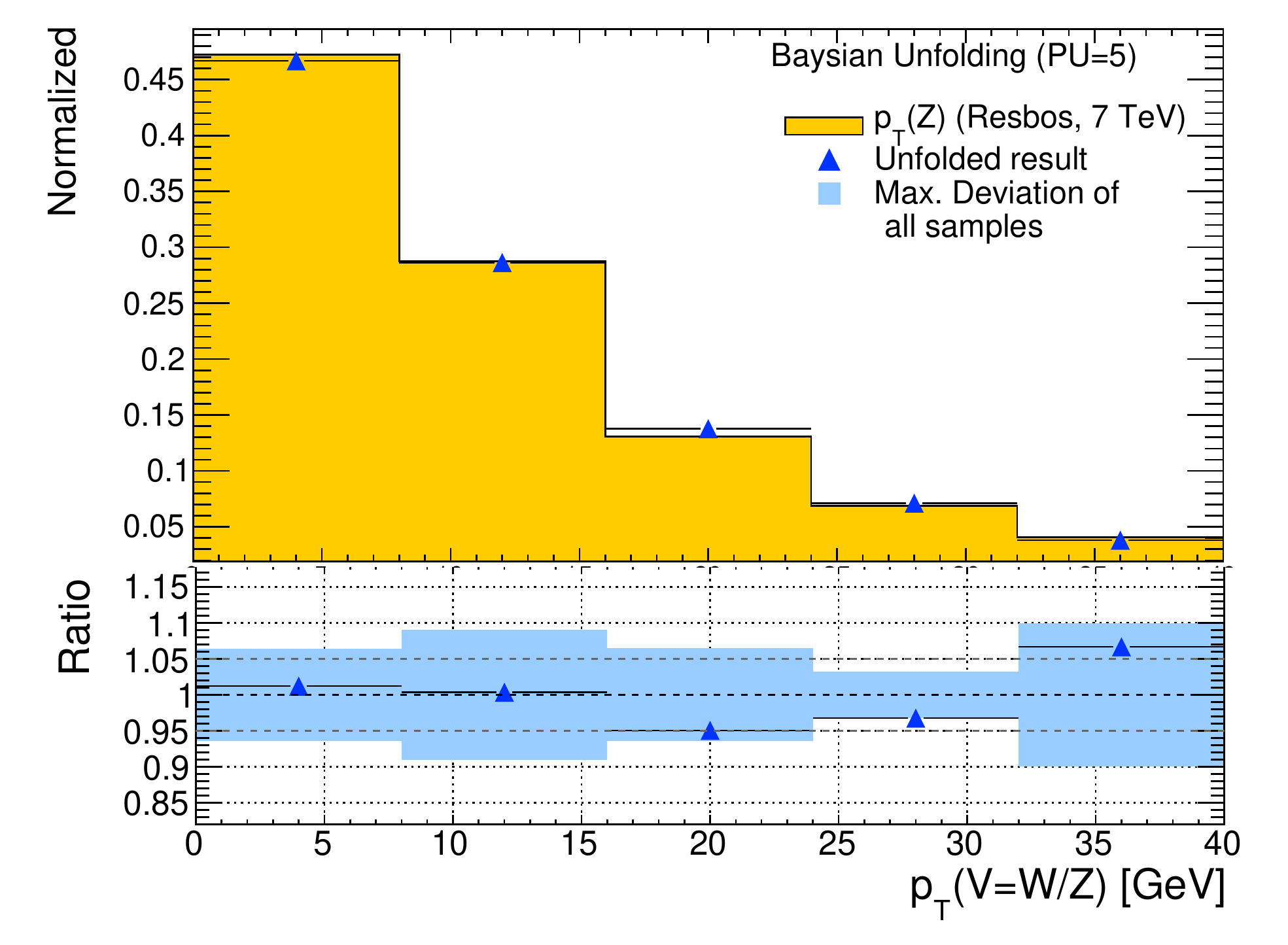}
        \includegraphics[width=0.49\textwidth]{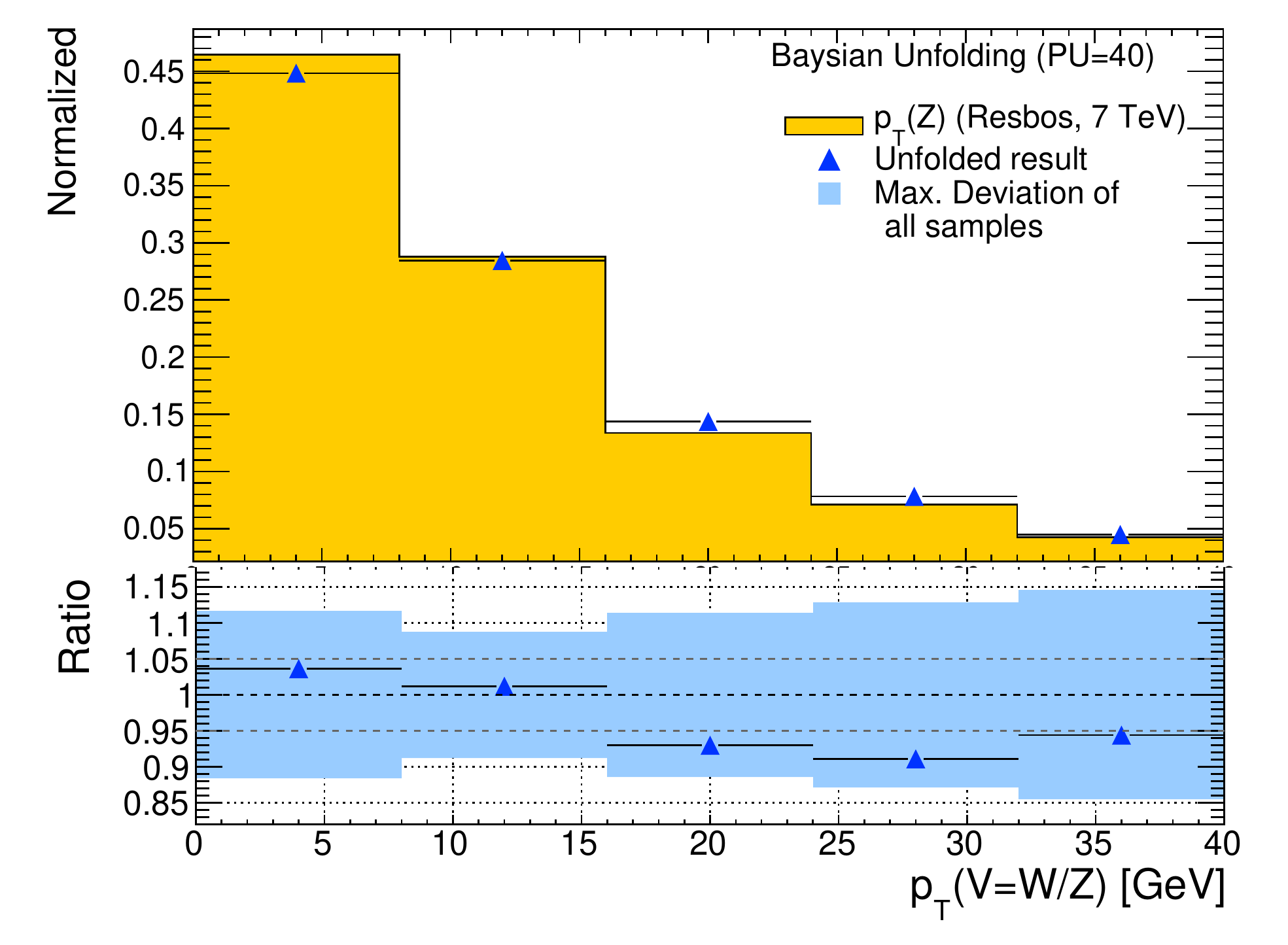}
        \caption{Comparison of the measured $p_T(Z)$ distribution and the corresponding unfolded distribution using the measured hadronic recoil at $\braket{\mu}=5$ and $\braket{\mu}=40$. The ratio plots also include the systematic uncertainty band due to the unfolding procedure. See text for further details. }
        \label{Fig:UnfoldedResults}
    \end{center}
\end{figure}

In a second step, the different pseudo distributions corresponding to the various MC predictions of Table \ref{DataTable} have been used as input and the unfolding was performed with three iterations\footnote{No significant differences have been observed when using five iterations during the Bayesian unfolding}. Each unfolded distribution was compared to the corresponding truth $p_T(V)$ spectrum. Similarly to the accuracy tests of the function description in Section \ref{Sec:accTests}, we defined an overall uncertainty band as the maximal observed deviation between the truth and unfolded distribution of all pseudo-data samples in each bin. In addition, we also used again the published $p_T(Z)$ distribution to reweight the simulated MC samples, leading to data-like hadronic recoil distributions. This was used as input for the Bayesian unfolding. The resulting unfolded distribution is then compared to the original published distribution for the two different $\braket{\mu}$ values in Figure \ref{Fig:UnfoldedResults}. Also shown is the uncertainties band, which is based on all available pseudo-data distributions. As expected, we see significantly better results for low $\mu$ values and larger binnings. However, the expected uncertainty is still in the order of 5\% for the 8 GeV binning and $\braket{\mu}=5$ and rises to more than $10\%$ for $\braket{\mu}=40$. 

\subsection{\label{sec:FunctionalFit}Expected Performance of the Functional Forward Folding Approach}

The performance of the functional fitting approach is tested analogously to the Bayesian unfolding as described in the previous section. For each pseudo-data hadronic recoil distribution, we apply the methodology that was introduced in Section \ref{sec:Met}. Equation \ref{eqnpara2} with fixed parameters $\rho$ and $\tau$ is used as a functional description of the transverse momentum distribution. Initial parameter values for the \textsc{Minuit}-based $\chi^2$ minimization are determined by a grid-scan over roughly 30 different starting values for each parameter\footnote{The \textsc{Simplex} minimization method was used, implemented in the \textsc{Root} package.}. The parameter set giving the lowest $\chi^2$ value is taken as the starting set for the \textsc{Minuit} routine, if it is not at the boundaries of the allowed parameter ranges \footnote{As a cross-check, we also performed a minimization with default starting parameters. For the vast majority of the cases, the pre-scanned values lead to better results.}. In the latter case, the fit is started with default values. 

\begin{figure}[tb]
    \begin{center}
        \includegraphics[width=0.49\textwidth]{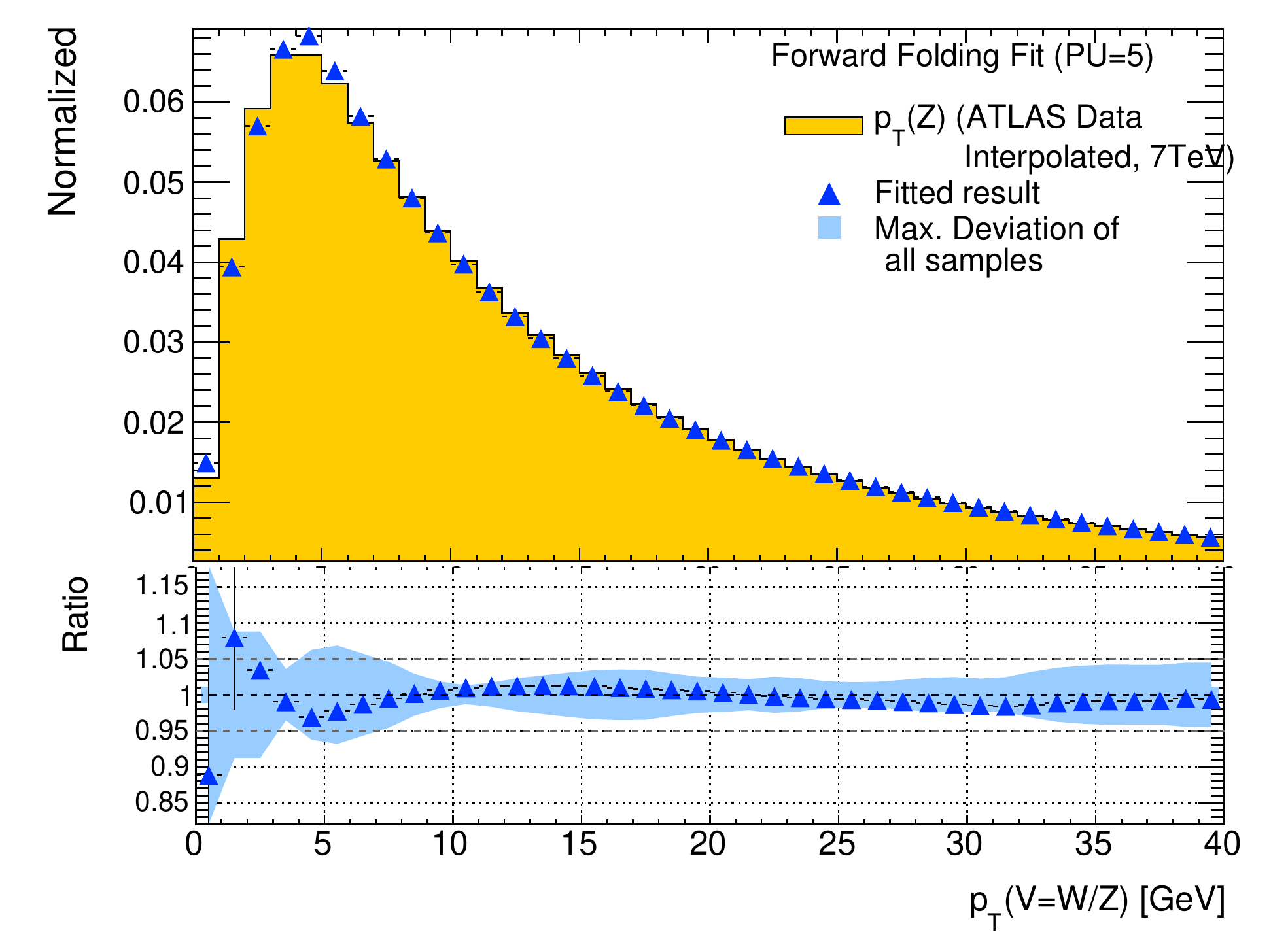}
        \includegraphics[width=0.49\textwidth]{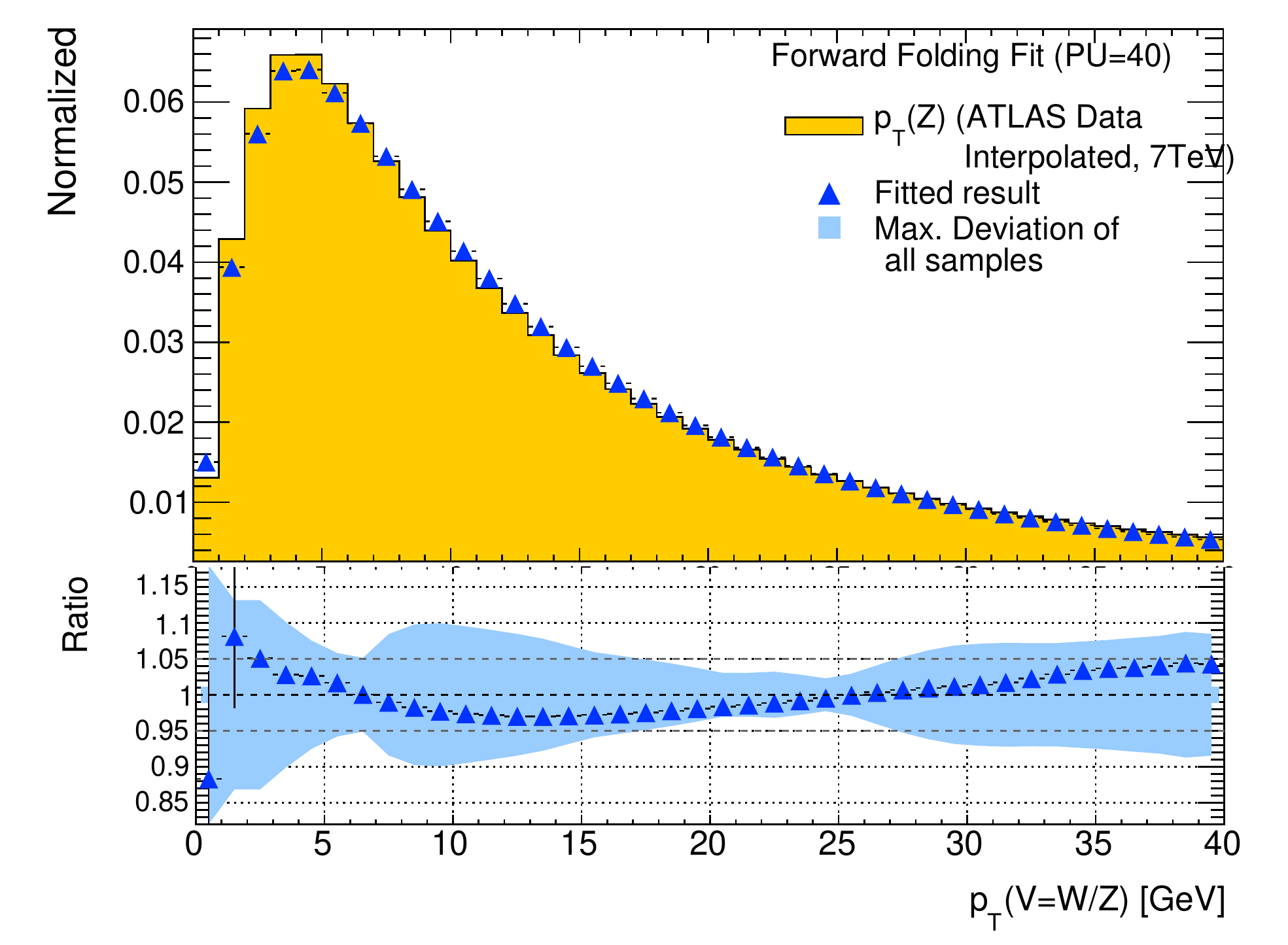}
        \caption{Comparison of the measured $p_T(Z)$ distribution and the corresponding unfolded distribution using the measured hadronic recoil at $\braket{\mu}=5$ and $\braket{\mu}=20$ using the functional fit approach. The ratio plots also includes the systematic uncertainty band due to the unfolding procedure. See text for further details. }
        \label{Fig:FunctionResults}
    \end{center}
\end{figure}

We determine the best matching parameter set for each pseudo-data set and compare the corresponding functional description to the original truth distribution. The uncertainty band of this method is then defined analogously to the Bayesian evaluation, i.e. we define the maximal deviation of the fitted functional description to the original truth distribution of all pseudo-datasets and each bin. As a reference example, we show the result when taking the measured $p_T(Z)$ distribution as input to define the pseudo-dataset. The results, including the uncertainty bands, are shown for two different pile-up settings in Figure \ref{Fig:FunctionResults}. The maximal deviation can be seen for small $p_T(V=W,Z)\lesssim3$\,GeV. The average uncertainty based on all pseudo-datasets is $2-4\%$ and $5-8\%$ for $\braket{\mu}=5$ and $\braket{\mu}=40$, respectively. As expected, the achievable precision is also reduced for higher pile-up scenarios when using the functional fitting approach, since the available information on $p_T(W)$ is more diluted. Studies for a pile-up scenario of $\braket{\mu}=20$ have been also performed, which resulted in a very similar precision as we observed for $\braket{\mu}=40$.

\begin{figure}[tb]
\begin{minipage}[t]{7.5cm}
	\centering
	\includegraphics[width=7.4cm]{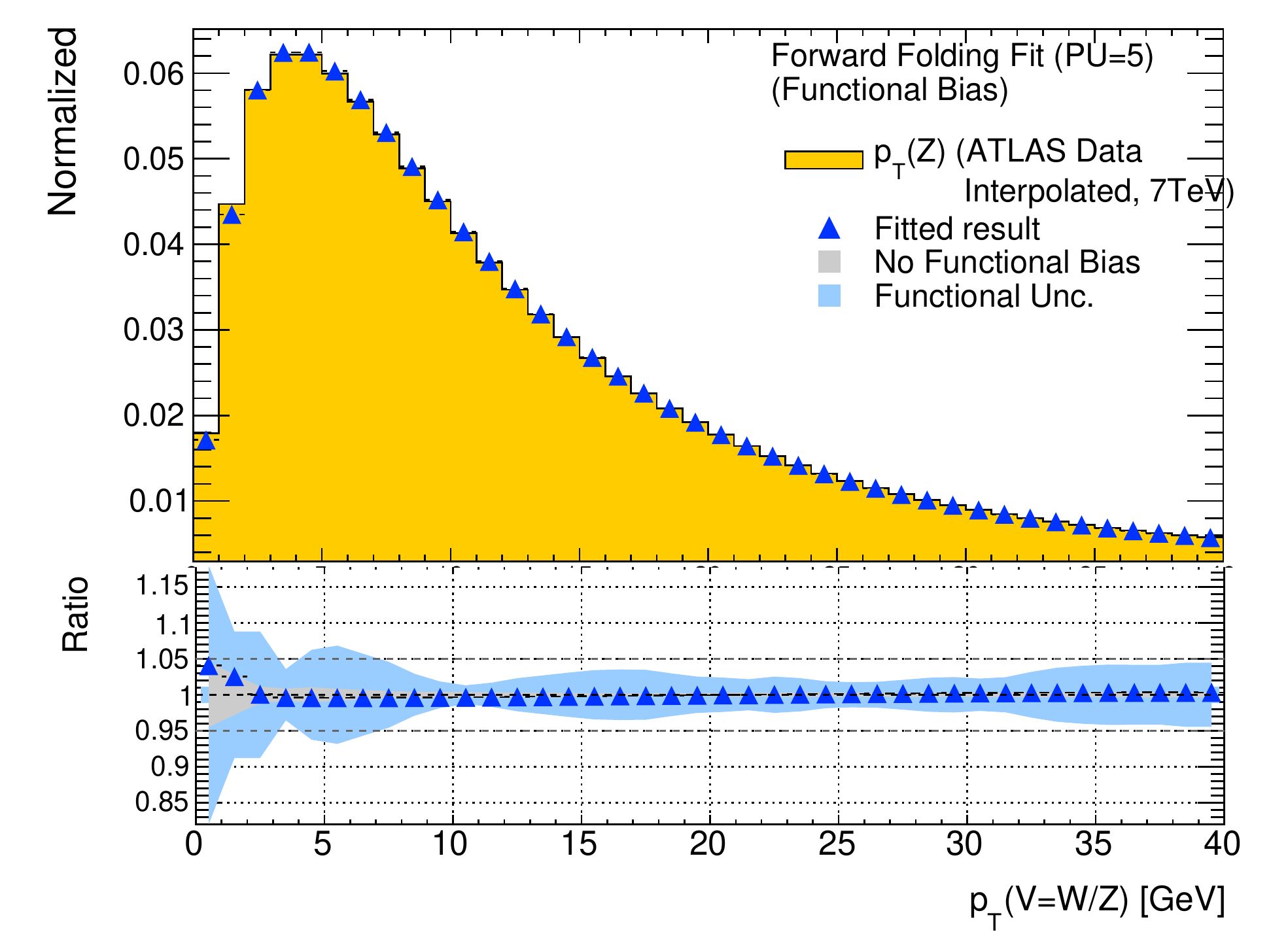}
	\caption{Comparison of the measured $p_T(Z)$ distribution, which was modified to be perfectly described by Equation \ref{eqnpara2} with fixed parameters $\rho$ and $\tau$ and the corresponding unfolded distribution using the measured hadronic recoil at $\braket{\mu}=5$ using the functional fit approach. The ratio plot also includes the systematic uncertainty band due to the nominal unfolding procedure (blue) and the uncertainty band, if the fitting function were to perfectly describe the truth distribution (gray).}
	\label{Fig:FuncBias}
\end{minipage}
\hfill
\hspace{0.1cm}
\begin{minipage}[t]{7.5cm}
	\centering
	\includegraphics[width=7.4cm]{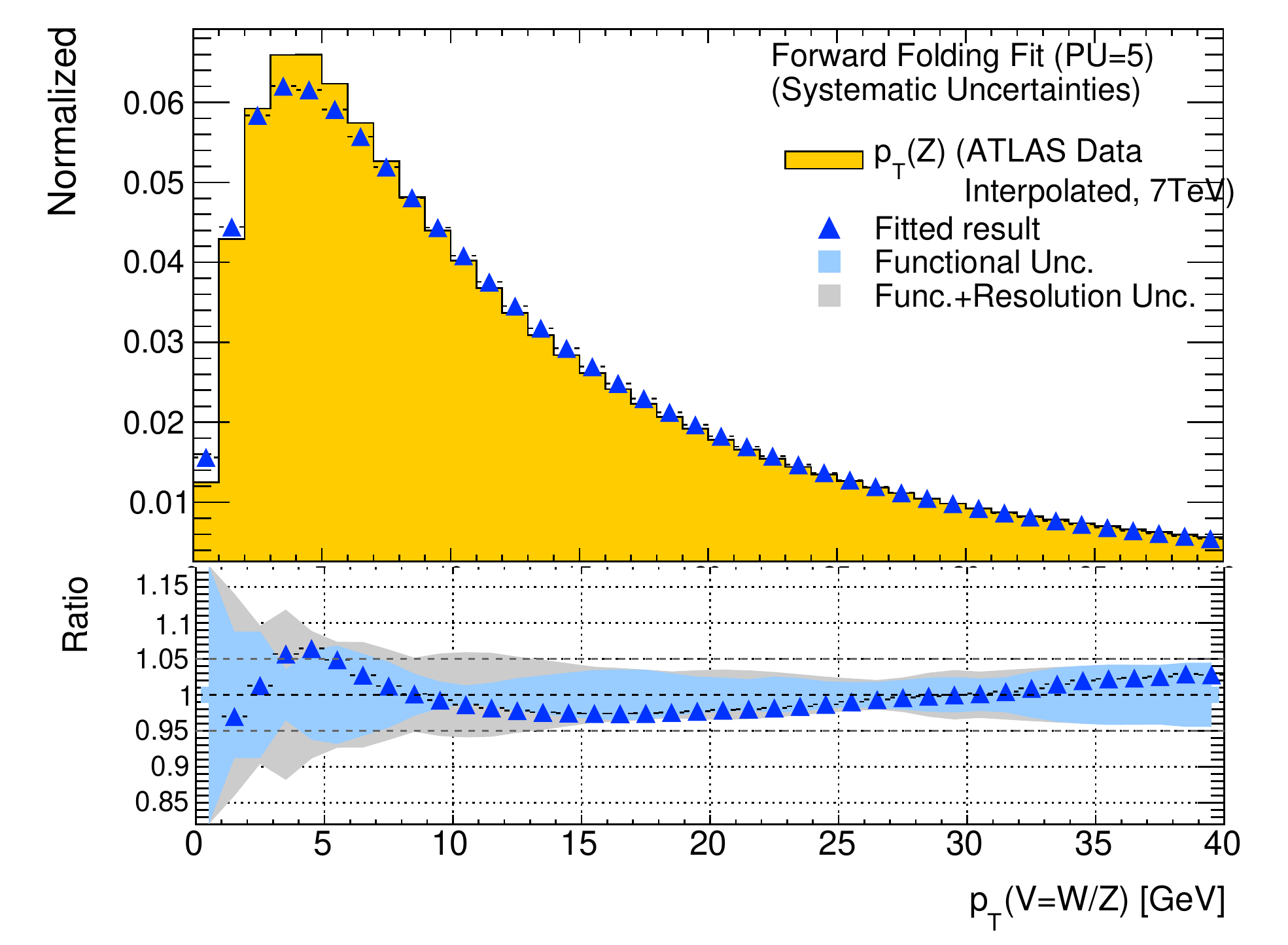}
	\caption{Comparison of the measured $p_T(Z)$ distribution and the corresponding unfolded distribution using the measured hadronic recoil at $\braket{\mu}=5$ with a systematic difference in the assumed hadronic recoil resolution of 5\% using the functional fit approach. The ratio plot also includes the systematic uncertainty band due to the nominal unfolding procedure (blue) and the uncertainty band including the additional uncertainty on the hadronic recoil resolution (gray).}
	\label{Fig:FuncSys}
\end{minipage}
\end{figure}

In order to verify the fitting approach and study the origin of the observed deviations, we modified each pseudo-data $p_T(V=W,Z)$ distribution such that it can be perfectly described by Equation \ref{eqnpara2} with fixed parameters $\rho$ and $\tau$. In this scenario, we expect that the fitting approach should find the correct $\chi^2$ minimum. The results are shown in Figure \ref{Fig:FuncBias}, where this assumption is confirmed. Hence the ability of a chosen functional form to describe the potential truth distribution is the limiting factor of the forward folding procedure. The methodology bias introduced by using the chosen parameterized form of $p_T(W)$ is similar to the functional bias of PDF fits, since here also a certain shape is assumed, which might differ from reality. It should be noted, that the systematic uncertainties of this technique can be also tested directly on data, by comparing the unfolded $p_T(Z)$ spectrum in Z boson events based on the full kinematics of decay leptons and the resulting functional form based on the hadronic recoil information.

Experimental systematic uncertainties, such as the knowledge of the resolution of the hadronic recoil measurement, will also impact the final result. In order to give a estimate of these effects, we have changed the hadronic recoil resolution relatively by $\approx5\%$ during the forward folding procedure, which mimics a difference between the assumed and the real detector description. The comparison of the unfolded ATLAS data set compared with the expected distribution using the stated difference in the hadronic recoil resolution is shown in Figure \ref{Fig:FuncSys}. The ratio  also indicates the systematic uncertainty band due to the nominal unfolding procedure and the uncertainty band including the additional uncertainty on the hadronic recoil resolution. While the expected uncertainties do not change for $p_T(W)>15\,$GeV, we see an increase to 5\% in the region of 8 to 15 GeV, as well as a modest increase in uncertainty for the very low $p_T$ region.

\subsection{\label{sec:Comp}Comparison of Results}

\begin{figure}[bt]
    \begin{center}
        \includegraphics[width=0.49\textwidth]{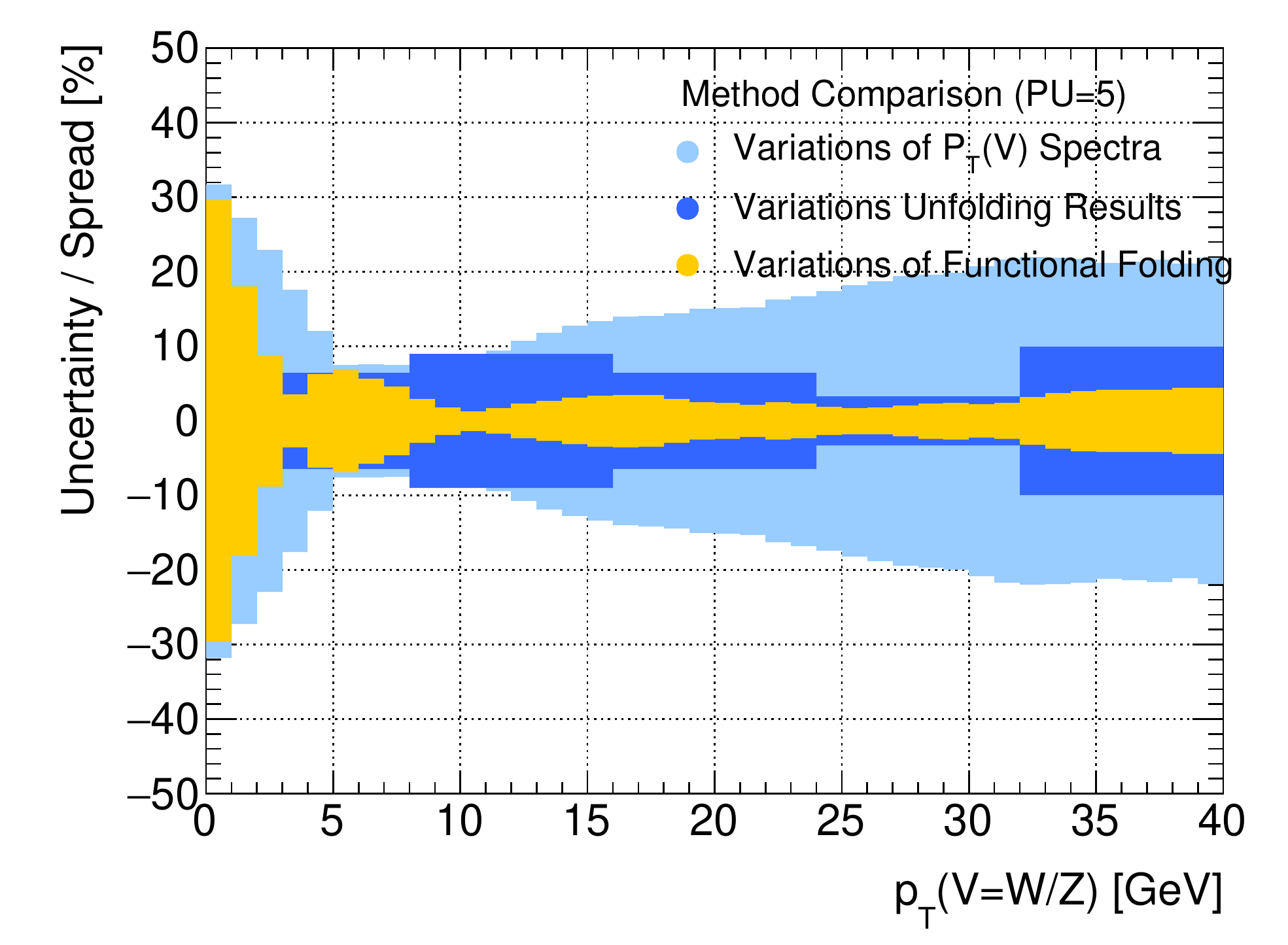}
        \includegraphics[width=0.49\textwidth]{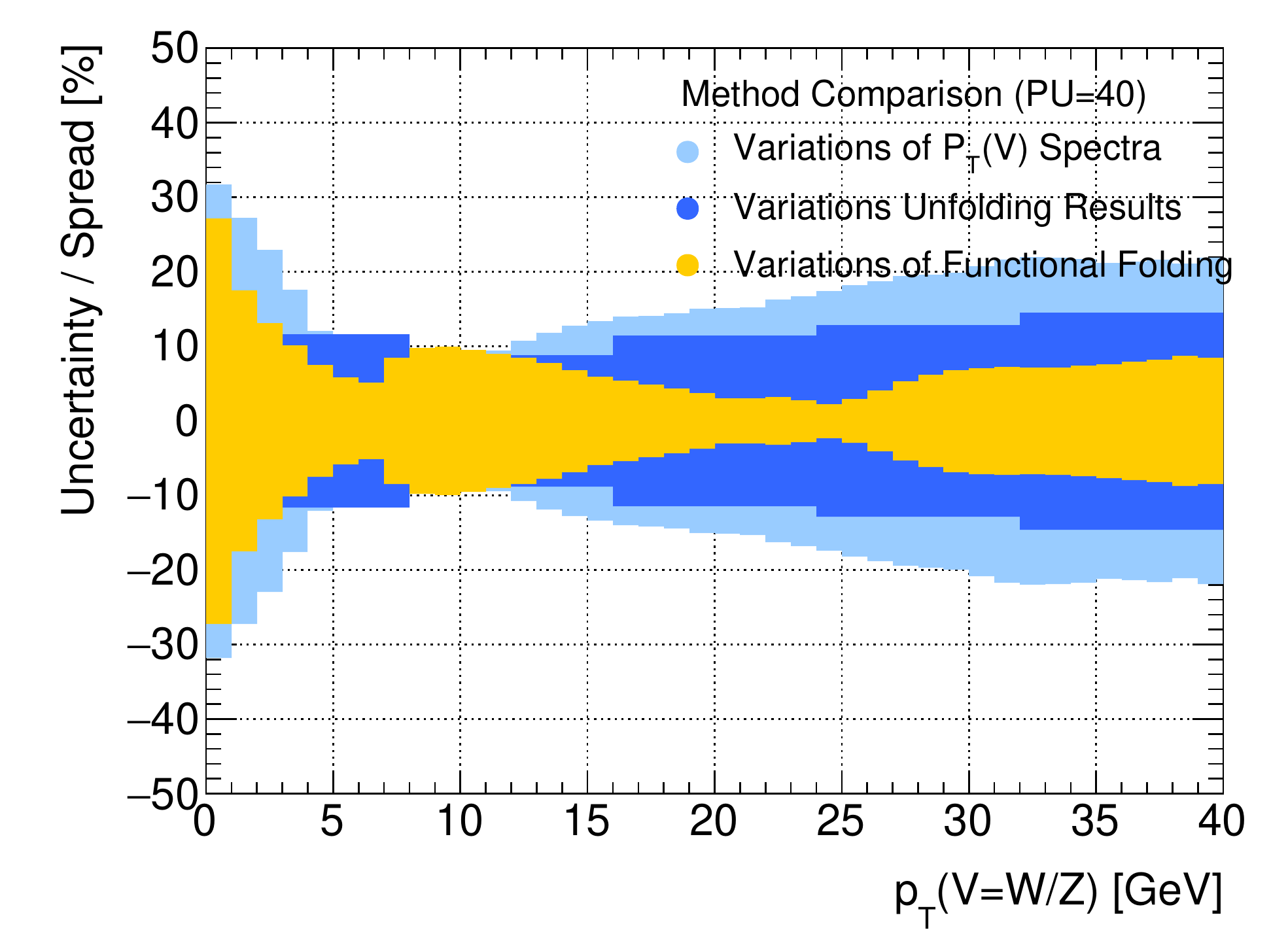}
        \caption{Comparison of the expected methodology uncertainties for the Bayesian unfolding technique and the functional fitting approach for two different pile-up scenarios.}
        \label{Fig:ResultComparison}
    \end{center}
\end{figure}

The Bayesian unfolding is limited by the large bin-to-bin migration effects for the hadronic recoil observable, making a relatively large binning necessary to ensure stable unfolded results\footnote{Further experimental uncertainties in the response matrix are not considered here, as they would impact the forward folding technique in a similar manner.}. The limitations of the functional fitting approach are due to the accuracy of the function description of the true boson $p_T$ distribution. A direct comparison of the expected systematic uncertainties due to the methodology is shown in Figure \ref{Fig:ResultComparison} for two pile-up scenarios. The maximal uncertainty of the functional forward fitting approach is significantly smaller over nearly the full range than the expected uncertainties in a Bayesian unfolding approach. Even more importantly, the functional fitting approach provides an unbinned description of the truth spectrum within the limitations of the fitting function. 

However, several aspects have to be considered. The presented methodology uncertainty of the functional fitting approach cannot be interpreted in a Gaussian way, but rather similarly to a \textit{range-fit} approach. Hence the truth distribution is expected to lie within the stated uncertainty band, which was already conservatively chosen, but no likelihood can be given. Moreover, no detector uncertainties have been considered, since they are expected to play a similar role for both approaches. 

\section{\label{sec:Sum}Summary and Conclusion}

In this paper we have presented a novel technique to infer the low $p_T(W)$ spectrum in hadronic collisions to high precision. The approach is based on a functional description of the $p_T(W)$ spectrum with only a few independent parameters. These parameters are varied such that the measured hadronic spectrum in data and MC prediction coincide within the associated uncertainties. 

It should be noted that this semi-empirical function description of $p_T(W)$ should be mainly seen as an illustrative example of the forward folding procedure. In future years, the functional description could be in principle directly taken from a theoretical prediction based on NNLO and NNLL calculations, which depends only on a small set of model parameters such as the strong coupling constant $\alpha_s$ and a resummation parameter $g$. The advantage of directly using physical parameters during the forward folding lies in the well defined resulting systematic uncertainties of these parameters. However, it is currently not yet possible to evaluate such a functional description following from QCD calculations, in a reasonable time, as a large number of generated events is required for each step in the fitting procedure. Therefore, we have presented a semi-empirical function with three independent parameters that describes a large variety of possible $p_T(W)$ spectra to high precision. With this function, the measured $p_T(W)$ spectrum is expected not to exceed a systematic uncertainty of typically $5\%$ up to transverse momenta of $40\,\GeV$. 

We observe a strong dependence on $\braket{\mu}$ of the expected precision for both Bayesian unfolding and the functional forward folding approach. Both approaches degrade significantly in precision when going from $\braket{\mu}=5$ to $\braket{\mu}=20$, but degrade only mildly when increasing the pile-up to $\braket{\mu}=40$. This behavior is expected, as the additional pile-up contribution should scale as $\sqrt{\braket{\mu}}$. 

While the tuning of model parameters of approximated QCD calculations will also be performed in future on the high precision $p_T(Z)$ data, the novel approach presented here allows a testing of these predictions using W boson data directly even in high pile-up environments. This would also allow one to estimate the uncertainty of the modeling of $p_T(W)$ on the W boson mass measurement in a rigorous way.

\section*{Acknowledgments}

We would like to thank Maarten Boonekamp, Aleksandra Dimitrievska and Nenad Vranjes, who started the initial studies and provided very helpful feedback throughout the work and on the final paper draft. In addition, we also would like to thank Marco Guzzi for valuable discussions and help with the \textsc{ResBos} generator. This work was supported by the Volkswagen Foundation and the German Research Foundation (DFG).

\bibliography{WPt}{}

\providecommand{\href}[2]{#2}\begingroup\raggedright\begin{thebibliography}{10}

\bibitem{Aad:2011fp}
{ATLAS} Collaboration, {\em {Measurement of the Transverse Momentum
  Distribution of $W$ Bosons in $pp$ Collisions at $\sqrt{s}=7$ TeV with the
  ATLAS Detector}\/},
  \href{http://dx.doi.org/10.1103/PhysRevD.85.012005}{Phys. Rev. {\bf D85}
  (2012)  012005},
\href{http://arxiv.org/abs/1108.6308}{{\tt arXiv:1108.6308 [hep-ex]}}.

\bibitem{D'Agostini:1994zf}
G.~D'Agostini, {\em {A Multidimensional unfolding method based on Bayes'
  theorem}\/},
\href{http://dx.doi.org/10.1016/0168-9002(95)00274-X}{Nucl. Instrum. Meth. {\bf
  A362} (1995)  487--498}.

\bibitem{Aad:2014xaa}
{ATLAS} Collaboration, {\em {Measurement of the $Z/\gamma^*$ boson transverse
  momentum distribution in $pp$ collisions at $\sqrt{s}$ = 7 TeV with the ATLAS
  detector}\/},  \href{http://dx.doi.org/10.1007/JHEP09(2014)145}{JHEP {\bf 09}
  (2014)  145},
\href{http://arxiv.org/abs/1406.3660}{{\tt arXiv:1406.3660 [hep-ex]}}.

\bibitem{James:1975dr}
F.~James and M.~Roos, {\em {Minuit: A System for Function Minimization and
  Analysis of the Parameter Errors and Correlations}\/},
\href{http://dx.doi.org/10.1016/0010-4655(75)90039-9}{Comput. Phys. Commun.
  {\bf 10} (1975)  343--367}.

\bibitem{Ladinsky:1993zn}
G.~A. Ladinsky and C.~P. Yuan, {\em {The Nonperturbative regime in QCD
  resummation for gauge boson production at hadron colliders}\/},
  \href{http://dx.doi.org/10.1103/PhysRevD.50.R4239}{Phys. Rev. {\bf D50}
  (1994)  4239},
\href{http://arxiv.org/abs/hep-ph/9311341}{{\tt arXiv:hep-ph/9311341
  [hep-ph]}}.

\bibitem{Balazs:1997xd}
C.~Balazs and C.~P. Yuan, {\em {Soft gluon effects on lepton pairs at hadron
  colliders}\/},  \href{http://dx.doi.org/10.1103/PhysRevD.56.5558}{Phys. Rev.
  {\bf D56} (1997)  5558--5583},
\href{http://arxiv.org/abs/hep-ph/9704258}{{\tt arXiv:hep-ph/9704258
  [hep-ph]}}.

\bibitem{Landry:2002ix}
F.~Landry, R.~Brock, P.~M. Nadolsky, and C.~P. Yuan, {\em {Tevatron Run-1 $Z$
  boson data and Collins-Soper-Sterman resummation formalism}\/},
  \href{http://dx.doi.org/10.1103/PhysRevD.67.073016}{Phys. Rev. {\bf D67}
  (2003)  073016},
\href{http://arxiv.org/abs/hep-ph/0212159}{{\tt arXiv:hep-ph/0212159
  [hep-ph]}}.

\bibitem{Boonekamp:2010ik}
M.~Boonekamp and M.~Schott, {\em {Z boson transverse momentum spectrum from the
  lepton angular distributions}\/},
  \href{http://dx.doi.org/10.1007/JHEP11(2010)153}{JHEP {\bf 11} (2010)  153},
\href{http://arxiv.org/abs/1002.1850}{{\tt arXiv:1002.1850 [hep-ex]}}.

\bibitem{Sjostrand:2007gs}
T.~Sjostrand, S.~Mrenna, and P.~Z. Skands, {\em {A Brief Introduction to PYTHIA
  8.1}\/},  \href{http://dx.doi.org/10.1016/j.cpc.2008.01.036}{Comput. Phys.
  Commun. {\bf 178} (2008)  852--867},
\href{http://arxiv.org/abs/0710.3820}{{\tt arXiv:0710.3820 [hep-ph]}}.

\bibitem{Alioli:2010xd}
S.~Alioli, P.~Nason, C.~Oleari, and E.~Re, {\em {A general framework for
  implementing NLO calculations in shower Monte Carlo programs: the POWHEG
  BOX}\/},  \href{http://dx.doi.org/10.1007/JHEP06(2010)043}{JHEP {\bf 06}
  (2010)  043},
\href{http://arxiv.org/abs/1002.2581}{{\tt arXiv:1002.2581 [hep-ph]}}.

\bibitem{Frixione:2007vw}
S.~Frixione, P.~Nason, and C.~Oleari, {\em {Matching NLO QCD computations with
  Parton Shower simulations: the POWHEG method}\/},
  \href{http://dx.doi.org/10.1088/1126-6708/2007/11/070}{JHEP {\bf 11} (2007)
  070},
\href{http://arxiv.org/abs/0709.2092}{{\tt arXiv:0709.2092 [hep-ph]}}.

\bibitem{Gleisberg:2008ta}
T.~Gleisberg, S.~Hoeche, F.~Krauss, M.~Schonherr, S.~Schumann, F.~Siegert, and
  J.~Winter, {\em {Event generation with SHERPA 1.1}\/},
  \href{http://dx.doi.org/10.1088/1126-6708/2009/02/007}{JHEP {\bf 02} (2009)
  007},
\href{http://arxiv.org/abs/0811.4622}{{\tt arXiv:0811.4622 [hep-ph]}}.

\bibitem{Gao:2013xoa}
J.~Gao, M.~Guzzi, J.~Huston, H.-L. Lai, Z.~Li, P.~Nadolsky, J.~Pumplin,
  D.~Stump, and C.~P. Yuan, {\em {CT10 next-to-next-to-leading order global
  analysis of QCD}\/},
  \href{http://dx.doi.org/10.1103/PhysRevD.89.033009}{Phys. Rev. {\bf D89}
  (2014) no.~3, 033009},
\href{http://arxiv.org/abs/1302.6246}{{\tt arXiv:1302.6246 [hep-ph]}}.

\bibitem{Watt:2012tq}
G.~Watt and R.~S. Thorne, {\em {Study of Monte Carlo approach to experimental
  uncertainty propagation with MSTW 2008 PDFs}\/},
  \href{http://dx.doi.org/10.1007/JHEP08(2012)052}{JHEP {\bf 08} (2012)  052},
\href{http://arxiv.org/abs/1205.4024}{{\tt arXiv:1205.4024 [hep-ph]}}.

\bibitem{Ovyn:2009tx}
S.~Ovyn, X.~Rouby, and V.~Lemaitre, {\em {DELPHES, a framework for fast
  simulation of a generic collider experiment}\/},
\href{http://arxiv.org/abs/0903.2225}{{\tt arXiv:0903.2225 [hep-ph]}}.

\end{thebibliography}\endgroup
\bibliographystyle{biblioint}

\end{document}